\documentstyle[floats,prl,aps,epsf]{revtex}
\draft % this command allows pacs numbers to be printed
\begin{document}

%%%%%%%%%%%%%%%%%%%%%%%%%%%%%%%%%%%%%%%%%%%%%%%%%%%%%%%%%%%%%%%%%%%%%%
%
%  Uncomment following four lines and one below for 2 column format
%  and figure insertions.
%
%\input epsf
%\renewcommand{\topfraction}{0.8}

%  This is the first line to be uncommented for 2 column format
\twocolumn[\hsize\textwidth\columnwidth\hsize\csname
@twocolumnfalse\endcsname
%
%%%%%%%%%%%%%%%%%%%%%%%%%%%%%%%%%%%%%%%%%%%%%%%%%%%%%%%%%%%%%%%%%%%%%%

\title{Gravitational wave extraction and 
	outer boundary conditions by perturbative matching}
\author{The Binary Black Hole Grand Challenge Alliance:} 
\author{A.~M.~Abrahams$^{\rm a,b,c}$, L.~Rezzolla$^{\rm a}$, 
	M.~E.~Rupright$^{\rm c}$,
	A.~Anderson$^{\rm c}$, P.~Anninos$^{\rm a}$, 
	T.~W.~Baumgarte$^{\rm a}$, N.~T.~Bishop$^{\rm d,l}$,
	S.~R.~Brandt$^{\rm a}$, J.~C.~Browne$^{\rm e}$, 
	K.~Camarda$^{\rm f}$, M.~W.~Choptuik$^{\rm e}$, 
	G.~B.~Cook$^{\rm g}$, C.~R.~Evans$^{\rm c}$, 
	L.~S.~Finn$^{\rm h}$, G.~Fox$^{\rm i}$, 
	R.~G\'omez$^{\rm j}$, T.~Haupt$^{\rm i}$, 
	M.~F.~Huq$^{\rm e}$, L.~E.~Kidder$^{\rm h}$, 
	S.~Klasky$^{\rm i}$, P.~Laguna$^{\rm f}$, 
	W.~Landry$^{\rm g}$, L.~Lehner$^{\rm j}$, 
	J.~Lenaghan$^{\rm c}$, R.~L.~Marsa$^{\rm e}$,
	J.~Masso$^{\rm a}$, R.~A.~Matzner$^{\rm e,k}$, 
	S.~Mitra$^{\rm e}$, P.~Papadopoulos$^{\rm f}$, 
	M.~Parashar$^{\rm e}$, F.~Saied$^{\rm a}$, 
	P.~E.~Saylor$^{\rm a}$, M.~A.~Scheel$^{\rm c}$,	
	E.~Seidel$^{\rm a}$, S.~L.~Shapiro$^{\rm a}$, 
	D.~Shoemaker$^{\rm e}$, L.~Smarr$^{\rm a}$, 
	B.~Szil\'agyi$^{\rm j}$, S.~A.~Teukolsky$^{\rm g}$, 
	M.~H.~P.~M.~van Putten$^{\rm g}$, P.~Walker$^{\rm a}$, 
	J.~Winicour$^{\rm j}$, J.~W.~York Jr$^{\rm c}$.}

\author{}
\address{$^{\rm a}$University of Illinois at
	           Urbana-Champaign, Urbana, IL 61801}
\address{$^{\rm b}$J.~P. Morgan, 23 Wall St., New York, NY 10260}
\address{$^{\rm c}$University of North Carolina, Chapel Hill, NC 27599}
\address{$^{\rm d}$University of South Africa, P.O. Box 392, 
	   	   Pretoria 0001, South Africa}
\address{$^{\rm e}$The University of Texas at Austin,  
   	           Austin, Texas 78712}
\address{$^{\rm f}$Penn State University, University Park, PA 16802}
\address{$^{\rm g}$Cornell University, Ithaca, NY 14853}
\address{$^{\rm h}$Northwestern University, Evanston, IL 60208}
\address{$^{\rm i}$Syracuse University, Syracuse, NY 13244-4100}
\address{$^{\rm j}$University of Pittsburgh, Pittsburgh, PA 15260}
\address{$^{\rm k}$Lead Principal Investigator}
\address{$^{\rm l}$Associate}

\maketitle

\begin{abstract}
We present a method for extracting gravitational radiation from a
three-dimensional numerical relativity simulation and, using the
extracted data, to provide outer boundary conditions.  The method
treats dynamical gravitational variables as nonspherical perturbations
of Schwarzschild geometry. We discuss a code which implements this
method and present results of tests which have been performed with a
three dimensional numerical relativity code.
\end{abstract}


% This is the other line to be uncommented for 2 column format
\vskip2pc]

	Numerical relativity represents the only currently viable
method for obtaining solutions to Einstein equations for highly
dynamical and strong field sources of gravitational radiation.  Using
these techniques to study coalescing black hole binaries is the
purpose of the multi-institutional Binary Black Hole ``Grand
Challenge'' {\it Alliance} effort\cite{bbh} which is presently
underway in the United States. This effort is also motivated by the
prospect of observations with the next generation of gravitational
wave detectors.

	In addition to tremendous demands on computational resources,
implementing the standard $3+1$ \cite{adm62,y79} formulation of
Einstein theory as a Cauchy problem \cite{cby80} is complicated
considerably by the necessity of imposing boundary conditions which
maintain numerical accuracy and the physical correctness of the
solution.  Both inner and outer boundary conditions have received
considerable attention.  Recent efforts on interior boundaries have
focused on the excision of the interior of the black hole from the
computational domain (see, for example, \cite{ss92}).  This paper will
concentrate on the problem of outer boundary conditions applied at a
finite radius around a source of gravitational waves.

	Proper boundary conditions on spacelike slices of
asymptotically flat spacetimes are {\it essential} for the accurate
computation of the gravitational waveforms produced in the strong
field region that represent the observationally relevant aspect of the
computation. Since it is not feasible to simulate on spacelike slices
out to arbitrarily large distances from the source, it is necessary to
extract gravitational waves comparatively near the strong field region
and to have boundary conditions that allow radiation to pass cleanly
off the mesh. If poor outgoing boundary conditions are imposed,
spurious radiation is produced which can contaminate the computed
gravitational waveform. Additionally, the outer boundary is usually
close enough to the isolated source that backscatter of radiation from
curvature is significant. This source of incoming radiation needs to
be built into the outer boundary conditions. An approach to the
extraction of gravitational wave information and the computation of
outer boundary conditions that exploits the matching of the interior
numerical solution with an exterior perturbative solution on spacelike
slices has been developed during the past decade and applied to a
number of different physical scenarios \cite{ae88,aetal92,ast95}.
Extension of these techniques to three dimensional (3D) simulations
has been one of the efforts of the Alliance.  A parallel development
is also underway in the Alliance that matches interior Cauchy
solutions to exterior solutions on characteristic hypersurfaces
\cite{ccm}.

	In this Letter we report the first successful application of
the perturbative matching approach to provide outer boundary
conditions for a 3D numerical relativity code. In this context, the
perturbative matching method should be viewed as a general ``module''
that can be coupled to any nonlinear ``interior'' 3D code. While the
latter solves the Einstein equations in the highly dynamical strong
field region, the former extracts gravitational wave data and imposes
outer boundary conditions. In the following we give an overview of the
method and present test results obtained from the evolution of linear
waves.

	Nonspherical perturbations of Schwarzschild geometry have been
studied for many years in the context of black hole perturbations
\cite{rwz,m74} and as a way to extract gauge-invariant information
about gravitational radiation \cite{ae88}. More recently Schwarzschild
perturbation theory has been found to be useful in studying the
late-time behavior of the coalescence of compact binaries in a
numerical simulation after the horizon has formed
\cite{pp94,ac94,ast95}.

	We base our treatment of Schwarzschild perturbation theory on
a recent hyperbolic formulation of Einstein field equations
\cite{aacby95}. A principal advantage of this approach is that we can
easily derive perturbative wave equations in terms of the standard
\hbox{$3+1$} variables, making it straightforward to match exterior
perturbative solutions to interior numerical ones. This method leads
to spatially gauge-invariant radial wave equations for each angular
mode \cite{aaelry} which we have used: {\sl a)} to extract
gravitational radiation from a 3D numerical relativity simulation,
{\sl b)} to evolve this information to large radius yielding an
approximate asymptotic waveform and {\sl c)} to provide outer boundary
conditions for such a simulation. One of the major advantages of the
perturbative method is that we have replaced the computationally
expensive 3D evolution of the gravitational waves via the nonlinear
Einstein equations with a set of 1D linear equations we can integrate
to high accuracy and minimal computational costs.

	We split the gravitational quantities of interest into
background and perturbed parts: the three-metric $g_{ij} = \tilde
g_{ij} + h_{ij}$, the extrinsic curvature $K_{ij} = \tilde{K}_{ij} +
\kappa_{ij}$, the lapse function $N = \tilde{N} + \alpha$ and the
shift vector $\beta^i = \tilde {\beta}^i + v^i$, where the tilde
denotes background quantities. We assume a Schwarzschild background,

\begin{eqnarray}
\tilde{g}_{ij} dx^i dx^j &=& \tilde{N}^{-2} dr^2 + 
	r^2 (d\theta^2 + \sin^2\theta d\phi^2) \ , \\
\tilde{N} & = & \left(1-\frac{2M}{r}\right)^{1/2} \ .
\end{eqnarray}
(It follows that $\tilde{K}_{ij}=0=\tilde {\beta}^i$). The perturbed
parts have arbitrary angular dependence.

	We use this approximation to linearize the hyperbolic
equations and, in particular, we find that the wave equation for
$K_{ij}$ reduces to a linear wave equation for $\kappa_{ij}$ involving
also the background lapse \cite{aaelry}. We separate the angular
dependence in this equation by expanding $\kappa_{ij}$ in terms of
tensor spherical harmonics $(\hat e_1)_{ij},\ldots,(\hat f_4)_{ij}$.
In the notation of \cite{m74}, we have

\begin{eqnarray}
\kappa_{ij} = & & a_\times (t,r) (\hat e_1)_{ij} + 
                r b_\times (t,r) (\hat e_2)_{ij} + \nonumber\\
     & & \tilde N^{-2} a_+ (t,r) (\hat f_2)_{ij} + 
                     r b_+ (t,r) (\hat f_1)_{ij} + \nonumber\\
               & & r^2 c_+ (t,r) (\hat f_3)_{ij} +
                   r^2 d_+ (t,r) (\hat f_4)_{ij} \ ,
\label{kappa_expand}
\end{eqnarray}
where $a_\times,\,b_\times$ are the odd-parity multipoles, while
$a_+,\break b_+,\,c_+,\,d_+$ are the even-parity ones. Note that we
have suppressed the angular indices $(\ell,m)$ of $(\hat
e_1)_{ij},\ldots,(\hat f_4)_{ij}$ and of $a_\times,\ldots,d_+$ over
which there is an implicit sum.

	In odd-parity, we take the wave equation for $\kappa_{r
\theta}$ and use the linearized momentum constraints, $\tilde\nabla_k
(\kappa^k_{\ i} - \delta^k_{\ i} \kappa^{j}_{\ j}) = 0$, to eliminate
the odd-parity amplitude ($b_\times$). In even-parity, we use the wave
equation for $\kappa_{r r}$ together with the wave equation obtained
for the trace $\kappa = \kappa^{j}_{\ j} = h(t,r) Y_{_{\ell m}}$ and
the linearized momentum constraints. In this way we eliminate $b_+$
and $c_+$ and obtain two coupled equations for $a_+$ and $h$. For
each $(\ell,m)$ mode, we therefore have one odd-parity equation

\vbox{
\begin{eqnarray}
\Biggl\{ \partial^2_t - \tilde{N}^4 \partial^2_r - 
	\frac{2}{r}\tilde{N}^2 \partial_r 
        - \frac{2 M}{r^3} \left(1 - \frac{3 M}{2 r} \right) + 
	\hskip 1.0truecm \nonumber \\
	\tilde{N}^2 \left[ \frac{\ell(\ell+1)}{r^2} - 
	\frac{6 M}{r^3} \right] \Biggr\} 
        (a_\times)_{_{\ell m}} = 0 \ ,
\label{oddwave}
\end{eqnarray} }
and two coupled even-parity equations,

\begin{eqnarray}
& & \hskip 0.0truecm \Biggl[ \partial^2_t - \tilde{N}^4 \partial^2_r - 
	\frac{6}{r}\tilde{N}^4 \partial_r
	+\tilde{N}^2 \frac{\ell(\ell+1)}{r^2} - \frac{6}{r^2} +
	\nonumber \\
	& &\frac{14M}{r^3}-\frac{3M^2}{r^4}
	\Biggr] (a_+)_{_{\ell m}} 
	+ \Biggl[\frac{4}{r} \tilde{N}^2 \left(1 -\frac{3M}{r}\right) 
	\partial_r +
	\nonumber\\ 
	& & \hskip 2.0truecm \frac{2}{r^2} 
	\left(1 - \frac{M}{r} - \frac{3M^2}{r^2}\right) 
	\Biggr] (h)_{_{\ell m}} = 0 \ , 
\label{evenwave1} 
\end{eqnarray}	

\begin{eqnarray}	
& & \hskip 0.0truecm 
	\Biggl[ \partial^2_t - \tilde{N}^4 \partial^2_r - 
	\frac{2}{r}\tilde{N}^2 \partial_r
	+ \tilde{N}^2 \frac{\ell(\ell+1)}{r^2} 
	+ \frac{2 M}{r^3} -
	\nonumber\\
	& & \hskip 0.6truecm
	\frac{7 M^2}{r^4} \Biggr] (h)_{_{\ell m}} 
	- \frac{2 M}{r^3} \left(3 - \frac{7 M}{r}\right) 
	(a_+)_{_{\ell m}} = 0 \ .
\label{evenwave2}
\end{eqnarray}
(The usual Regge-Wheeler and Zerilli equations can be obtained by a
more complete analysis \cite{aaelry}.) 

	We now turn to implementation and tests of this technique.
Consider a numerical relativity simulation which evolves Einstein
equations in either the standard $3+1$ (ADM) or hyperbolic form on a
``interior'' 3D grid. (No restrictions need be put on the choice of
the 3D coordinate system.) In the tests described here, we have used
the ADM 3D interior code of the Alliance \cite{adm_code} to evolve
linear, $\ell = 2,\;m=0$, Teukolsky waves \cite{t82}. The interior 3D
grid uses (topologically) Cartesian coordinates and up to $(129)^3$
gridpoints. In the results shown, the initial data consists of a
metric formed from the sum of the background Cartesian metric and a
Gaussian envelope of amplitude $A=10^{-6}$, width $b=1$ and of a
vanishing extrinsic curvature. The ADM evolution is undertaken on a
cubical grid of extent $\{x,y,z\} = \pm \;4$, with unit lapse and zero
shift.

\begin{figure}[t]
\epsfxsize=7.5cm
\begin{center}
\leavevmode
\epsffile{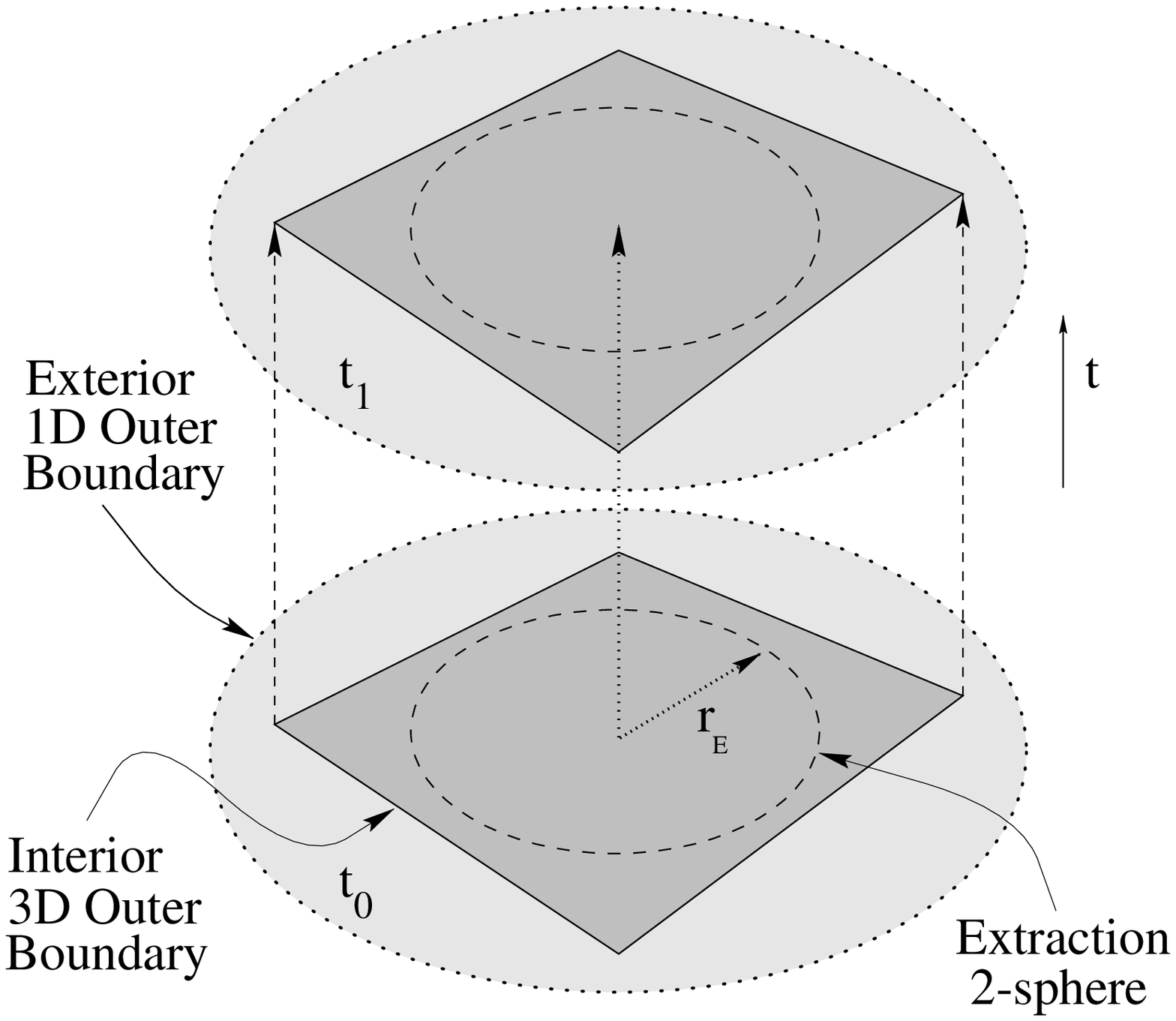}
\end{center}
\caption[fig1]{\label{fig1} 
Location of the different outer boundaries and of the extraction 
2-sphere for two successive timeslices. The dark shaded region shows
the spatial domain over which the 3D nonlinear equation are solved.}
\end{figure}

	During each time step, the procedure for extracting radiation
and imposing outer boundary conditions proceeds in three steps: {\sl
(1)} {\it extraction} of the independent amplitudes and their time
derivatives on a 2-sphere of radius $r_{_E}$ inscribed in the grid,
{\sl (2)} {\it evolution} of the radial wave equations
(\ref{oddwave})--(\ref{evenwave2}) out to large radii using the
extracted amplitudes to construct inner boundary values on the
``exterior'' 1D grid and {\sl (3)} {\it reconstruction} and {\it
``injection''} of $K_{ij}$ and $\partial_t K_{ij}$ at specified
gridpoints at or near the boundary of the interior grid. The last step
uses the momentum constraints and the inverse of the transformation
employed in step {\sl (1)}. In Figure \ref{fig1}, we show the
schematic location of the different outer boundaries and of the
extraction 2-sphere on two successive spacelike time slices at $t=t_0$
and $t=t_1$.

	The first step involves interpolating $K_{ij}$ and $\partial_t
K_{ij}$ onto the 2-sphere and evaluating their projections onto the
appropriate conjugate tensor spherical harmonics to eliminate the
angular dependence and obtain the corresponding multipoles. The
precise location of the 2-sphere within the interior 3D grid depends
on the specific case under investigation. In general, for the
perturbative matching to be justified, we require that the extraction
surface be located in a region where the gravitational field is close
to Schwarzschild. One must be careful that the numerical errors
introduced by the interior evolution do not dominate the exterior
solution. (Further details will be presented elsewhere \cite{arr97}.)

\begin{figure}[t]
\epsfxsize=8cm 
\begin{center}
\leavevmode
\epsffile{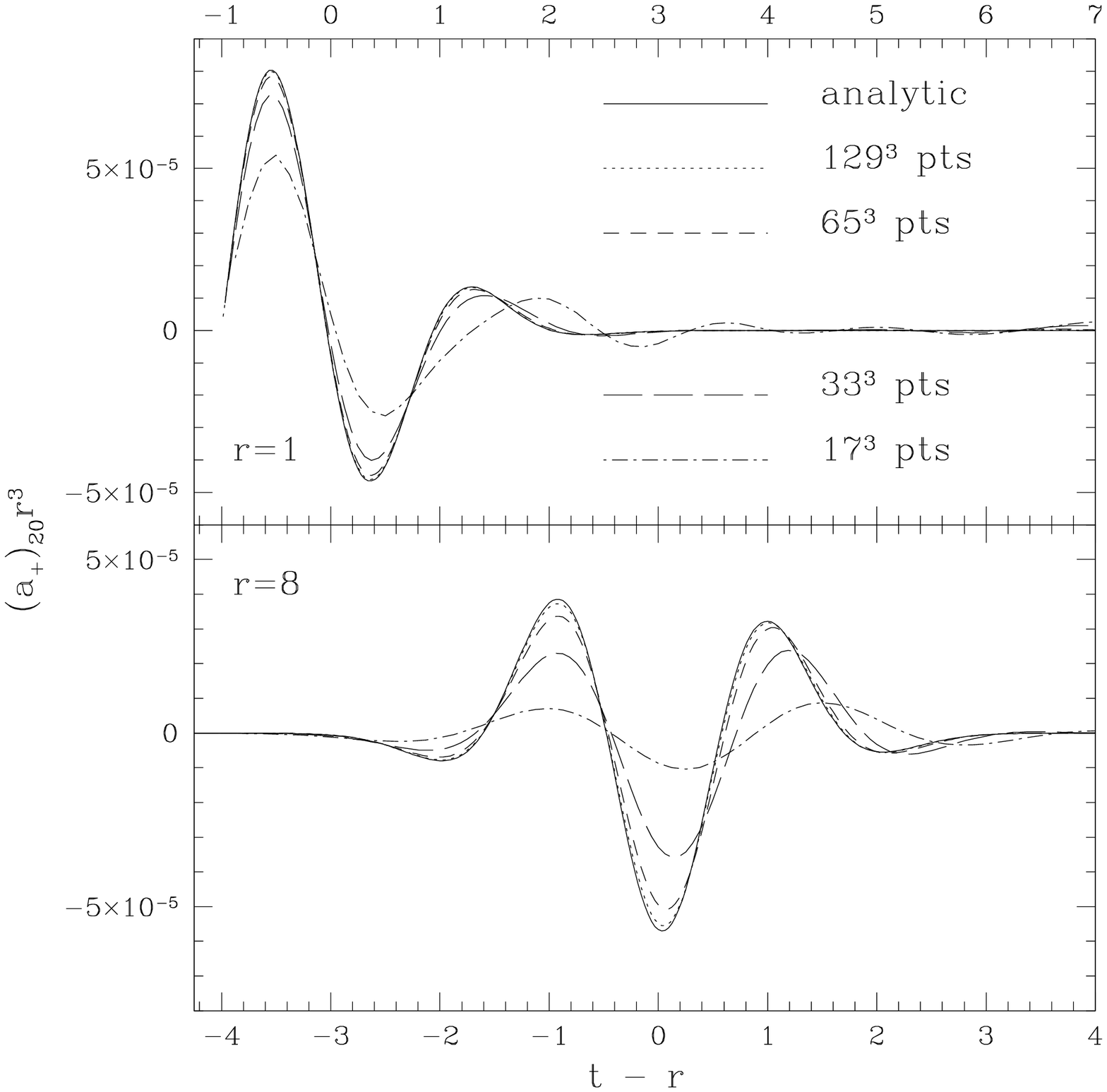} 
\end{center}
\caption[fig2]{\label{fig2} 
Convergence to the analytic solution of the extracted ($r=1$) and
evolved ($r=8$) multipole $(a_+)_{20}$. The amplitude is scaled by
$r^3$ to compensate for the radial fall-off.}
\end{figure}

	The second step entails evolving equations
(\ref{oddwave})--(\ref{evenwave2}). In the test presented here, we
consider a flat background with $M=0$. The initial data for the 1D
exterior grid is set consistently with the interior 3D initial
data. During each time integration of the 1D exterior grid, the
extracted multipoles are used as inner boundary values and standard
Sommerfeld outgoing wave conditions are imposed at the outer boundary
(e.g., at $r=30$). The top diagram of Fig. \ref{fig2} shows the only
relevant multipole for this test \cite{note_1}, extracted from the
interior 3D grid at a 2-sphere of radius $r_{_E}=1$ \cite{note_2}; the
lower diagram shows its evolved waveform at a radius $r = 8$.
Different curves refer to different resolutions of the interior 3D
grid and show the convergence to the analytic solution.

\begin{figure}[t]
\epsfxsize=7.5cm 
\begin{center}
\leavevmode
\epsffile{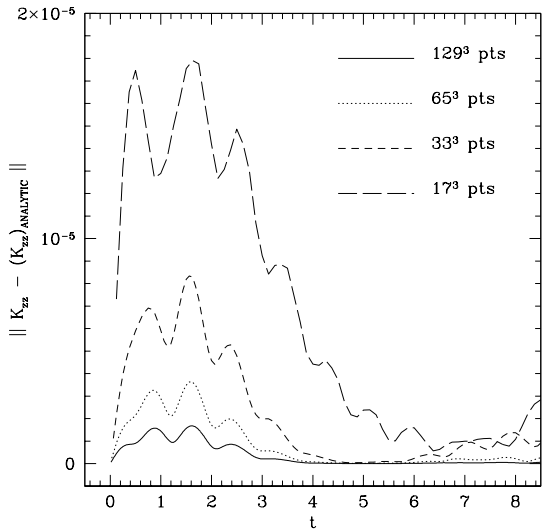}
\end{center}
\caption[fig3]{\label{fig3} 
Convergence to zero of the $L_2$ norm of the error of $K_{zz}$
integrated over the 3D outer boundary surface.}
\end{figure}

	The third and final step consists of computing outer boundary
values for the interior 3D grid, and this step can proceed in one of
two ways. In the first method, the values of $K_{ij}$ reconstructed
from the exterior 1D data are injected as Dirichlet data at all the
boundary points of the interior 3D grid. (No restrictions are put on
the shape of the 2-surface where the data are injected.)  For a code
using a hyperbolic formulation, $\partial_t K_{ij}$ can also be
provided at the boundary. Moreover, the evolution equation of the
three-metric can be integrated using $K_{ij}$ at the boundary so
Dirichlet data for $g_{ij}$ need not be provided there. The second
method relates, at the 3D outer boundary, the null derivatives of the
extrinsic curvature obtained from the interior grid ($K_{ij}$) and
from the perturbative module ($\kappa_{ij}$, since background
extrinsic curvature is assumed to be zero):

\begin{equation}
\label{ps}
{\partial \over \partial t} \left(K_{ij} - \kappa_{ij} \right) +
{\partial \over \partial r} \left(K_{ij} - \kappa_{ij} \right) +
{2 \over r} \left(K_{ij} - \kappa_{ij} \right) =0 
\end{equation}
This method resembles a Sommerfeld condition but is more general since
it can be used in regions where the radiation is not dominated by the
asymptotic outgoing behavior. Moreover it takes into account arbitrary
angular dependence, as well as the effects of a Schwarzschild black
hole background.  Experimentation with the two methods has shown that
this ``perturbative Sommerfeld'' approach is very accurate and
generally more stable than the direct injection of Dirichlet data.
Figure \ref{fig3} shows convergence to zero of the $L_2$ norm of the
difference between one component of $K_{ij}$ (namely $K_{zz}$) and its
analytic value, integrated over the entire outer boundary of the
interior 3D grid.

\begin{figure}
\epsfxsize=7.5cm
\begin{center}
\leavevmode
\epsffile{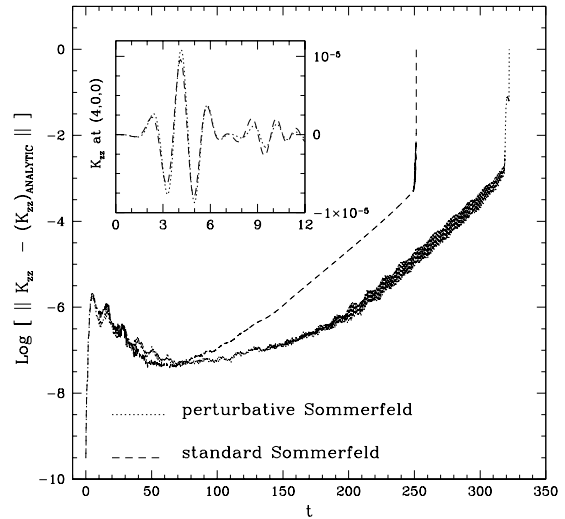} 
\end{center}
\caption[fig4]{\label{fig4} 
Long-term evolution of the $L_2$ norm of the error in $K_{zz}$ at the
3D outer boundary surface. The inset shows $K_{zz}$ at the outer
boundary along the $x-$axis. The interior grid has $(33)^3$ gridpoints
and the code is stopped when the norm is unity.}
\end{figure}

	An important property of {\it any} outer boundary
implementation is that it should allow the outgoing radiation to
escape freely to infinity. In practice, however, a finite
discretization always produces a certain amount of reflection at the
outer boundary and this could drive instabilities which grow
exponentially.  We have compared the long-term stability obtained with
the perturbative outer boundary module with that of the best
``alternative'' outer boundary conditions we have tried: namely a
standard outgoing Sommerfeld condition. Fig. \ref{fig4} shows the
results of this comparison both on a long time scale and on a shorter
time scale. (The calculations were performed using a very coarse
resolution.) As shown in the inset (first echoes at $t \sim 7$), the
amount of reflection produced with the perturbative method is smaller
than that produced by the Sommerfeld condition.  Moreover, the use of
perturbative boundary conditions delays the onset of exponential error
growth and allows for a much longer evolution (see main figure).
Increasing the interior resolution and the number of $(\ell, m)$ modes
used we can further prolong the running time. Using $(49)^3$ interior
gridpoints, we are able to evolve the code for more than $50$ crossing
times. In strong contrast, when a standard Sommerfeld condition is
used, increased interior resolution results in a shorter running time
\cite{arr97}.

	This work was supported by the NSF Binary Black Hole Grand
Challenge Grant Nos. NSF PHY 93--18152, NSF PHY 93--10083, ASC
93--18152 (ARPA supplemented). Computations were performed at NPAC
(Syracuse University) and at NCSA (University of Illinois at
Urbana-Champaign).

% References


\begin{references}
\bibitem{note_1}Other $(\ell, m)$ multipoles are also computed and 
		used in the decomposition; but, given the initial 
		data chosen, their amplitudes are
		several orders of magnitude smaller \cite{arr97}.
\bibitem{note_2}On a flat background spacetime, the position
		of the extraction 2-sphere is arbitrary.
		%We have here chosen a location which maximizes 
		%the amplitude of the projected multipoles.
		Similar results
		have been obtained also for 
		$r_{_E}=0.5,1.5,2,2.5,3,3.5$ \cite{arr97}.
\end{references}
\end{document}